\documentclass[
    aps,
    prapplied,
    reprint,
    floatfix,
    superscriptaddress
]{revtex4-2}

\usepackage{graphicx}
\usepackage{siunitx}
\usepackage{upgreek}
\usepackage{xcolor} 
\usepackage{comment}

\newcommand{\kecom}[1]{{\color{magenta} Kevin: ``#1''}}

\begin{document}


\title{Strain-overlap and surface-confinement engineering in AlScN/YIG magnetoacoustic devices} 



\author{Kevin K{\"u}nstle}
\email{kuenstle@rptu.de}
\homepage[]{https://physik.rptu.de/ags/ag-weiler}
\affiliation{Fachbereich Physik and Landesforschungszentrum OPTIMAS, Rheinland-Pf{\"a}lzische Technische Universit{\"a}t Kaiserslautern-Landau, 67663 Kaiserslautern, Germany}

\author{Kaya Gauch}
\affiliation{Fachbereich Physik and Landesforschungszentrum OPTIMAS, Rheinland-Pf{\"a}lzische Technische Universit{\"a}t Kaiserslautern-Landau, 67663 Kaiserslautern, Germany}

\author{Agn{\.e} {\v{Z}}ukauskait{\.e}}
\affiliation{Fraunhofer Institute for Electron Beam and Plasma Technology FEP, 01277 Dresden, Germany}
\affiliation{Institute of Solid State Electronics, Technische Universität Dresden,
01062 Dresden, Germany}

\author{Stephan Barth}
\affiliation{Fraunhofer Institute for Electron
Beam and Plasma Technology FEP, 01277 Dresden, Germany}

\author{Mathias Weiler}
\affiliation{Fachbereich Physik and Landesforschungszentrum OPTIMAS, Rheinland-Pf{\"a}lzische Technische Universit{\"a}t Kaiserslautern-Landau, 67663 Kaiserslautern, Germany}



\date{\today}

\begin{abstract}
We establish AlScN/YIG/GGG as an integrated piezoelectric-magnetic heterostructure for electrically generating surface acoustic waves (SAWs) and coupling them to spin waves in monocrystalline YIG. Combining electrical transmission measurements of interdigital-transducer delay lines with finite-element simulations, we identify the excited modes as Rayleigh-type SAWs whose surface confinement varies strongly with AlScN thickness. The resulting changes in the strain field within YIG directly modify the magnetoelastic coupling. Using the simulated complex strain components and an effective-driving-field model, we calculate the relative SAW attenuation for several AlScN thicknesses and wavelengths. The calculations qualitatively reproduce the measurements: at all investigated wavelengths, thinner AlScN layers yield stronger relative attenuation. Experimentally, the attenuation approaches \qty{70}{\decibel\per\milli\meter} for a \qty{54.5}{\nano\meter}-thick AlScN layer at a wavelength of \qty{1.2}{\micro\meter}. However, the strongest relative attenuation does not maximize the detectable signal. Instead, the absolute field-induced transmission change peaks at an AlScN thickness of approximately \qty{100}{\nano\meter}, with the precise optimum depending on the acoustic wavelength. These results identify the AlScN thickness as a key design parameter for balancing magnetoelastic attenuation and pure SAW transmission in AlScN/YIG devices.

\end{abstract}

\maketitle 



\section{Introduction}\label{sec:intro}
Elastic strain provides a versatile route for exciting and controlling magnetization dynamics, enabling the investigation of a broad range of magnetoelastic phenomena \cite{Puebla2022-ja,Yang2021-yz}. Surface acoustic waves (SAWs) are particularly attractive for this purpose because they generate coherent strain fields at microwave frequencies and non-zero wave vectors \cite{royer1999elastic,Krenner2026-wi}. Their interaction with spin system has enabled phenomena ranging from strain-induced magnetization switching \cite{Thevenard2016-ou, Davis2010} and the manipulation of magnetic textures \cite{Yang2024-rd, Schwenke2025-gg} to strongly nonreciprocal acoustic transmission \cite{Kus2023-wh}. The latter can arise from the direction-dependent overlap between the elliptically polarized SAW and the resonantly precessing magnetization, often described as a helicity mismatch effect \cite{Kus2021-lr}. When the magnetoelastic coupling exceeds the relevant acoustic and magnetic loss rates, SAWs and spin waves (SWs) can furthermore form coherently hybridized magnon–phonon states \cite{Hwang2024-ks, Kuenstle2025}, which can even be driven into nonlinearity \cite{Alekseev2025-dk, Kunstle2026-bn}. Realizing and controlling these effects requires both low magnetic damping and a sufficiently large overlap between the acoustic strain field and the magnetic layer.

Due to its exceptionally low Gilbert damping, Yttrium iron garnet (YIG) is a key material for magnetoacoustic and magnonic devices \cite{CHEREPANOV199381, Serga2010YIGMagnonics, Kunz2025-mf}. Because YIG itself is not piezoelectric, an additional material is required to efficiently generate acoustic waves electrically. Previous approaches include the direct growth of polycrystalline YIG on piezoelectric $\mathrm{LiNbO_3}$ \cite{Holzmann2025-mv}, the transfer of YIG slabs onto $\mathrm{LiNbO_3}$ \cite{Ba2025-dz}, and the deposition of a piezoelectric thin film onto high-quality epitaxial YIG \cite{Ryburn2025-fq,Hatanaka2025-nb,Rivard2025-je}. Polycrystalline YIG grown on $\mathrm{LiNbO_3}$ generally does not retain the crystalline quality and low damping characteristic of epitaxial YIG on GGG, whereas transfer-based approaches require additional fabrication and bonding steps and may introduce variability at the YIG/piezoelectric interface. By contrast, depositing the piezoelectric material directly onto epitaxial YIG avoids removing the magnetic film from its growth substrate and enables the excitation of propagating SAWs in continuous, high-quality YIG films. This direct-overlayer approach has been employed in zinc oxide (ZnO)/YIG and aluminum nitride (AlN)/YIG heterostructures \cite{Ryburn2025-fq,Hatanaka2025-nb,Rivard2025-je}.

In this direct-overlayer geometry, the thickness of the piezoelectric film relative to the acoustic wavelength governs the dispersion, polarization, and depth profile of the acoustic eigenmodes \cite{FARNELL197235} and thereby the strain field within the magnetic layer. Importantly, the magnetoelastic driving field depends on the relevant strain components inside YIG rather than on the total acoustic displacement or surface amplitude alone. Consequently, a layer geometry that optimizes electromechanical SAW generation does not necessarily maximize magnetoelastic energy transfer.

Scandium-alloyed aluminum nitride (AlScN) offers a promising alternative to ZnO and AlN for magnetoacoustic integration. Sc incorporation enhances the piezoelectric response of AlN while retaining compatibility with reactive sputter deposition and integrated high-frequency acoustic devices \cite{Zukauskaite2012-yt,Kurz2019-nz,Urban2021-pg}. However, how the AlScN thickness and SAW wavelength jointly determine acoustic-mode formation, strain transfer into YIG, and the resulting magnetoelastic response within the AlScN/YIG/GGG stack has not yet been systematically established.

To address these questions, we combine finite-element (FEM) simulations with magnetic-field-dependent SAW-transmission measurements on delay-line devices spanning a range of AlScN thicknesses and acoustic wavelengths. The simulations yield the acoustic eigenmodes, their surface confinement, and the relative amplitudes and phases of the strain components within YIG, from which we calculate the corresponding magnetoelastic driving fields and expected SW absorption. The calculations predict, and the measurements confirm, that the SAW transmission normalized to the uncoupled response of each device is reduced with as the AlScN layer becomes thinner, corresponding to enhanced relative magnetoelastic attenuation. By contrast, the absolute magnetically induced transmission change is maximized at an intermediate AlScN thickness depending on the acoustic wavelength. For our parameters, this thickness is approximately \qty{100}{\nano\meter}. Our results establish AlScN/YIG as a viable magnetoacoustic platform and provide practical guidelines for selecting the piezoelectric-layer thickness and acoustic wavelength in future device designs.

\section{Material- and device design}\label{sec:device}

To investigate magnetoelastic coupling in AlScN/YIG heterostructures, we use a \qty{100}{\nano\meter}-thick liquid-phase epitaxy-grown YIG film on a GGG substrate \cite{Dubs2017LPEYIG}. The AlScN layers with different thicknesses were deposited onto the YIG by reactive pulsed DC magnetron sputtering. The depositions were done using a Double Ring Magnetron DRM 400 of Fraunhofer FEP with two AlSc alloy targets \cite{Barth2022}. For this study, we choose a Sc concentration of \qty{35}{\percent} in the target material. The resulting film composition ($\mathrm{Al_{0.65}Sc_{0.35}N}$) enables high piezoelectric response, while still not severely compromising the crystal structure and therefore enhancing propagation losses \cite{Zywitzki2017-nu}. Argon and nitrogen (5N) were used as inert and reactive gas, respectively. Stabilization of the process was done by a closed-loop reactive gas control. No active heating was used, and the film deposition was done at room temperature. The chamber base pressure was $\qty{2e-7}{\milli\bar}$. The investigated AlScN thicknesses are summarized in Tab.~\ref{tab:thicknesses}.

\begin{table}[ht]
    \caption{Nominal thicknesses of the investigated AlScN layers.}
    \label{tab:thicknesses}
    \begin{ruledtabular}
        \begin{tabular}{ccccc}
            \multicolumn{5}{c}{$t_{\mathrm{AlScN}}$ (nm)} \\ 
            \hline
            54.5 & 94 & 202 & 300 & 395
        \end{tabular}
    \end{ruledtabular}
\end{table}

\begin{figure}[ht]
\includegraphics{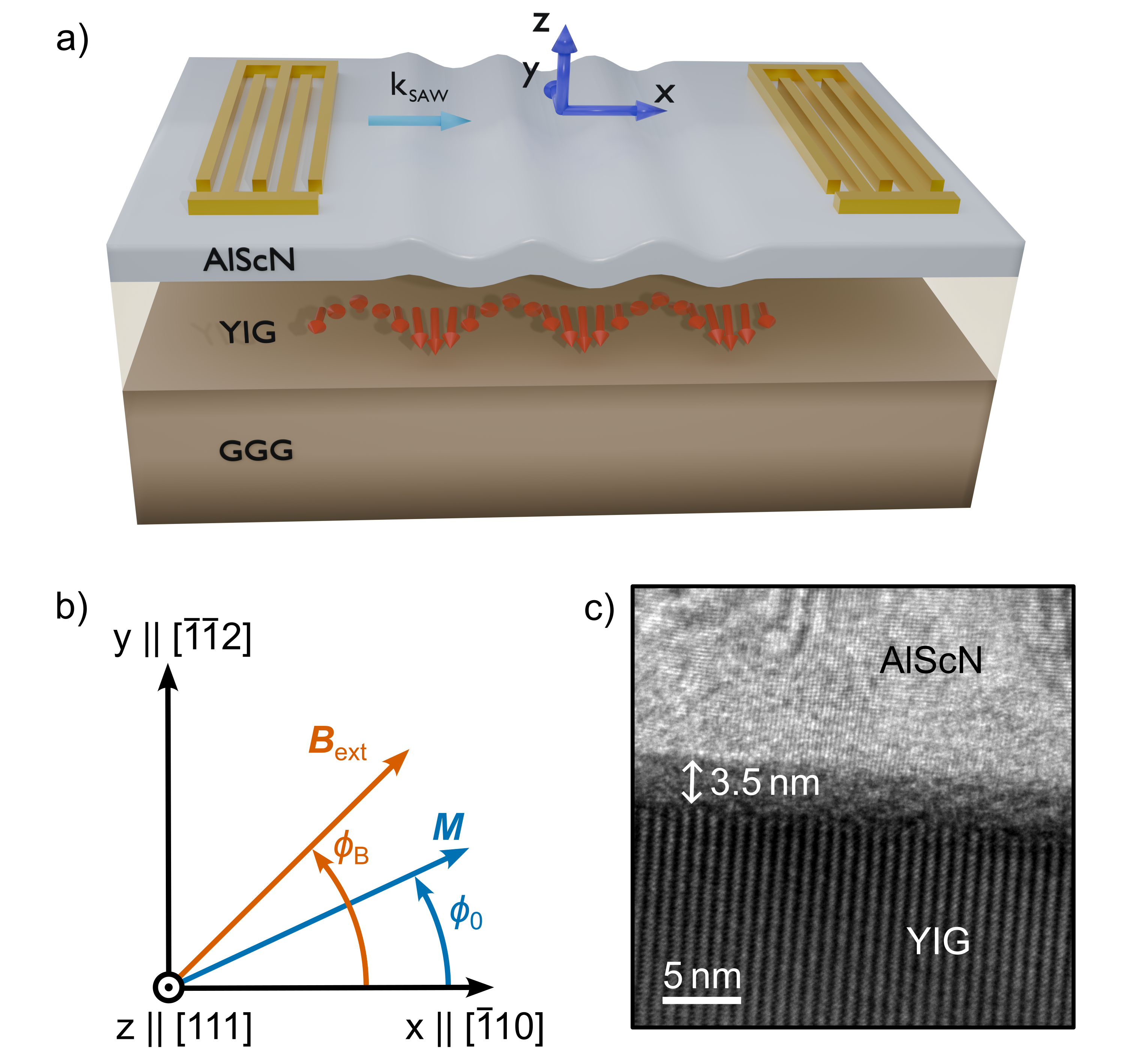}
\caption{\label{fig:schematic} a) Schematic depiction of the material stack, microstructures, and SAW propagation direction in the device coordinate system. b) Definition of the device coordinate system relative to the crystallographic axes of the YIG film, together with relevant angles. c) Transmission electron microscope image of the AlScN/YIG, showing a \qty{3.5}{\nano\meter}-thick amorphous interlayer.}
\end{figure}

A schematic of the investigated devices is shown in Fig.~\ref{fig:schematic}~a). Each device comprises a SAW delay line consisting of an input IDT for SAW excitation and an opposing output IDT for detection, both patterned on top of the piezoelectric AlScN layer by electron-beam lithography. The IDT electrodes consist of a \qty{5}{\nano\meter} Ti/\qty{30}{\nano\meter} Au bilayer. Three IDT designs were fabricated with nominal fundamental acoustic wavelengths of \qty{1.2}{\micro\meter}, \qty{2.0}{\micro\meter} and \qty{2.8}{\micro\meter}. For all devices, the center-to-center separation between the transmitting and receiving IDTs is \qty{500}{\micro\meter}. Further details of the fabrication process and device dimensions are provided in Sec.~S1 of the supplement.

The dynamic strain field of the SAW extends through the AlScN layer into the underlying YIG, where it couples to the magnetization dynamics through the magnetoelastic interaction \cite{Dreher2012-lw, Kunz2025-mf}. For a SAW with frequency $f_{\mathrm{SAW}}$ and wave vector $\mathbf{k}_{\mathrm{SAW}}$, the external magnetic field can be used to tune the SW dispersion into resonance. Resonant coupling occurs when $f_{\mathrm{SW}}(\mathbf{k}_{\mathrm{SAW}},B_{\mathrm{ext}})= f_{\mathrm{SAW}}$, thereby fulfilling both frequency and wave-vector matching. Under this condition, the SAW excites a SW in YIG and transfers energy from the acoustic wave to the magnetic system, resulting in an attenuation of the transmitted SAW signal.

The coordinate system and measurement geometry are defined in Fig.~\ref{fig:schematic}~b). The SAW propagates along the $x$ direction, parallel to the crystallographic $[\bar{1}10]$ axis of the YIG film. The in-plane angles $\phi_{\mathrm{B}}$ and $\phi_0$ denote the angles between the SAW propagation direction and the in-plane external magnetic field $\mathbf{B}_{\mathrm{ext}}$ and equilibrium magnetization $\mathbf{M}_0$, respectively. These angles determine the angular dependence of the magnetoelastic interaction considered in Sec.~\ref{sec:angle}.

A cross-sectional transmission electron microscopy image of the AlScN/YIG interface is shown in Fig.~\ref{fig:schematic}~c). The observed lattice contrast indicates that both AlScN and YIG are crystalline and separated by an amorphous interfacial layer with a thickness of approximately \qty{3.5}{\nano\meter}. XRD measurements confirm that the AlScN crystallizes in the wurtzite phase and is c-axis oriented, without any additional crystalographic phases or abnormally oriented grains \cite{Osterlund2021-hk} (see supplement Sec.~S2). This interfacial structure contrasts with previously studied ZnO/YIG heterostructures, for which an amorphous ZnO "dead layer" with a thickness of approximately \qty{200}{\nano\meter} was considered in Ref. \cite{Ryburn2025-fq}. The formation of crystalline AlScN within only a few nanometers of the YIG interface therefore represents a potential advantage for thin-film magnetoacoustic devices, as it facilitates the use of thinner piezoelectric layers. As shown below, reducing the AlScN thickness increases the acoustic-strain overlap with the YIG and can thereby enhance the magnetoelastic interaction.

\section{Thickness-dependent acoustic modes and surface confinement}\label{sec:thickness}

\begin{figure}
\centering
\includegraphics[width=\columnwidth]{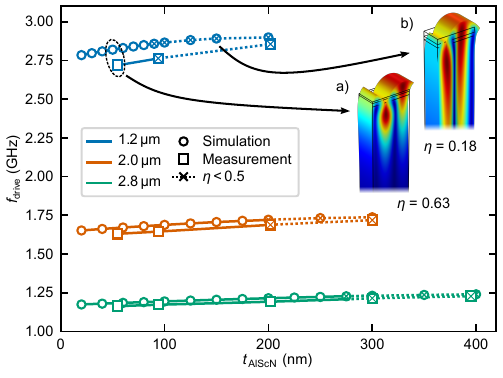}
\caption{\label{fig:freq} Measured peak frequencies (squares) and FEM-simulated eigenfrequencies (circles) of the acoustic mode as a function of the AlScN thickness $t_\mathrm{AlScN}$ for $\lambda=\qty{1.2}{\micro\meter}$, $\qty{2.0}{\micro\meter}$, and $\qty{2.8}{\micro\meter}$. Lines serve as guides to the eye. Crosses identify thickness-wavelength combinations for which the simulated surface-confinement parameter satisfies $\eta <0.5$. The insets show the normalized displacement magnitude for $\lambda=\qty{1.2}{\micro\meter}$ at a) $t_\mathrm{AlScN}=\qty{50}{\nano\meter}$ and b) $t_\mathrm{AlScN}=\qty{150}{\nano\meter}$, for which $\eta_\mathrm{surf}=0.63$ and $0.18$, respectively. The displayed simulation domain extends \qty{3.2}{\micro\meter} into the GGG substrate.}
\end{figure}

To verify SAW propagation in the delay lines, we utilize a vector network analyzer to electrically measure the complex transmission coefficient $S_{21}$ \cite{Pozar2012MicrowaveEngineering} of devices designed to excite acoustic wavelengths of $\lambda=\qty{1.2}{\micro\meter}$, $\qty{2.0}{\micro\meter}$, and $\qty{2.8}{\micro\meter}$, as set by the respective IDT periodicities. The delayed acoustic response is isolated from electromagnetic crosstalk using time-domain gating \cite{Pupalaikis_Doshi_2013}. For each device, the central SAW frequency $f_\mathrm{drive}$ is identified as the frequency at which the gated transmission magnitude reaches its maximum. The extracted frequencies are shown as squares in Fig.~\ref{fig:freq}.

We compare the measured frequencies with finite-element eigenfrequency simulations of the AlScN/YIG/GGG heterostructure performed using COMSOL Multiphysics\textsuperscript{\textregistered}~\cite{COMSOLMultiphysics}. The model describes the AlScN layer within linear piezoelectricity and includes anisotropic linear elasticity throughout the heterostructure. The full elastic and piezoelectric tensors are transformed into the device coordinate system shown in Fig.~\ref{fig:magelast}~b). Details of the simulation geometry, boundary conditions, and material parameters are provided in the supplement Sec.~3. The resulting eigenfrequencies, shown as circles in Fig.~\ref{fig:freq}, reproduce both the absolute frequency scale and its systematic increase with $t_\mathrm{AlScN}$. The associated eigenmodes exhibit predominantly sagittal-plane character, with strain fields dominated by the longitudinal component $\varepsilon_{xx}$ and the vertical-shear component $\varepsilon_{xz}$ in approximate phase quadrature. Together with the surface confinement quantified below, this strain composition identifies the modes as Rayleigh-like waves. An exemplary comparison between the simulated strain composition and that inferred from the angle-dependent measurements is presented in Table~\ref{tab:strain_comparison} of Sec.~\ref{sec:angle}, while the characteristic angular dependence of the magnetoelastic coupling provides further experimental support for the simulated strain symmetry.

Because the phase velocity is given by $v_\mathrm{p}=f\lambda$, the observed increase in frequency with $t_\mathrm{AlScN}$ at fixed $\lambda$ corresponds to a rise in the effective phase velocity. For the adopted directional material parameters, the relevant bulk-wave velocities of $\mathrm{Al}_{0.65}\mathrm{Sc}_{0.35}\mathrm{N}$ exceed those of YIG and GGG, placing the heterostructure in the fast-on-slow regime (see supplement Sec.~4). Accordingly, within the investigated thickness range, the increasing influence of the faster AlScN overlayer on the acoustic dispersion shifts the modes to higher frequencies as $t_\mathrm{AlScN}$ increases. The agreement between experiment and simulation is closest for $\lambda=\qty{2.8}{\micro\meter}$, whereas larger deviations are observed for $\lambda=\qty{1.2}{\micro\meter}$. At a fixed $t_\mathrm{AlScN}$, decreasing $\lambda$ increases the normalized layer thickness $t_\mathrm{AlScN}/\lambda$, making the mode frequency more sensitive to the elastic and piezoelectric properties of the AlScN layer as well as to lithographic and electrode-related uncertainties. The complete AlScN material tensors used in the simulations were taken from theoretical calculations for a Sc fraction of \qty{35}{\percent} \cite{Urban2021-pg}. Differences between these theoretical parameters and the effective properties of the sputtered film may therefore contribute to the residual deviations between the measured and simulated frequencies.

From the simulations, we can quantify the surface confinement using the fraction of the time-averaged mechanical energy \cite{Schuetz2015} contained within a depth of $\lambda/2$ from the top surface:

\begin{equation}
    \label{eq:eta}
    \eta
    =
    \frac{\displaystyle\int_{\Omega_{\lambda/2}}
    \left\langle w_\mathrm{kin}\right\rangle\,\mathrm{d}V}
    {\displaystyle\int_{\Omega}
    \left\langle w_\mathrm{kin}\right\rangle\,\mathrm{d}V},
    \qquad
    \left\langle w_\mathrm{kin}\right\rangle
    \propto
    \rho(\mathbf r)\omega^2
    \hat{\mathbf u}^{\dagger}\hat{\mathbf u}.
\end{equation}

Here, $\Omega$ denotes the complete simulated solid domain, $\Omega_{\lambda/2}$ denotes the part of this domain within $\lambda/2$ of the top surface, $\rho(\mathbf r)$ is the local mass density, and $\hat{\mathbf u}$ is the complex displacement phasor. In this work, we adopt $\eta>0.5$ as an criterion for classifying a mode as strongly surface confined. This threshold is used consistently throughout the figures to provide a direct visual indication of the AlScN thickness beyond which the modes no longer exhibit strong surface confinement. For example, in Fig.~\ref{fig:freq}, the thickness-wavelength combinations for which the simulated modes  are not surface confined, characterized by $\eta<0.5$, are marked by crosses.

The insets in Fig.~\ref{fig:freq}~a,b) illustrate the evolution of the mode profile for $\lambda=\qty{1.2}{\micro\meter}$. Increasing $t_\mathrm{AlScN}$ from \qty{50}{\nano\meter} to \qty{150}{\nano\meter} is accompanied by a pronounced extension of the displacement field into the underlying layers, reducing $\eta$ from $0.63$ to $0.18$. Across all three wavelengths, the AlScN thickness at which $\eta$ falls below $0.5$ increases with $\lambda$, demonstrating that the confinement is governed primarily by the normalized thickness $t_\mathrm{AlScN}/\lambda$ rather than by $t_\mathrm{AlScN}$ alone.

The progressive loss of surface confinement is consistent with the substrate-radiation cutoff expected for fast-on-slow layered structures \cite{FARNELL197235, Lefeuvre1998-ct, Lefeuvre1998-ht}. When $v_\mathrm{p}$ exceeds the lowest symmetry-allowed bulk-wave threshold of the GGG substrate for the conserved in-plane wave vector, the corresponding substrate partial wave changes from evanescent to propagating, thereby opening a bulk-radiation channel \cite{Darinskii1997LeakyWaves,Laude2004SAWSlowness}. The bound Rayleigh-like branch can consequently continue as a leaky (pseudo-surface) wave rather than necessarily terminate \cite{Naumenko1998LeakyLayered,Every2018FanoLeakySAW}. Consequently, selecting $t_\mathrm{AlScN}$ requires balancing at least three wavelength-dependent considerations: the electromechanical coupling coefficient $K^2$, which contributes to the efficiency with which the IDTs generate and detect the acoustic mode \cite{Smith1969-mf, Wang2014-xz}; the preservation of surface confinement to avoid acoustic radiation into the GGG substrate; and the spatial overlap of the relevant complex strain components with the YIG layer \cite{Yamamoto2022-sk}, which determines the magnetoelastic driving field and thereby the strength of the SAW-SW interaction. The optimum AlScN thickness is therefore wavelength dependent and depends on whether the device is primarily optimized for acoustic transduction or magnetoelastic coupling.

\section{Angle dependence of magnetoelastic coupling}\label{sec:angle}

Having established that the AlScN/YIG/GGG delay lines support surface-localized acoustic modes, we next investigate their resonant coupling to SWs. The angular dependence of this interaction also provides an experimental consistency check of the simulated acoustic mode shapes. We first consider the sample with $t_{\mathrm{AlScN}}=\qty{54.5}{\nano\meter}$ and the delay line designed for $\lambda=\qty{1.2}{\micro\meter}$. In this device, the strongly surface-localized mode occurs at $f_{\mathrm{drive}}=\qty{2.720}{\giga\hertz}$.

\begin{figure*}[t]
\includegraphics{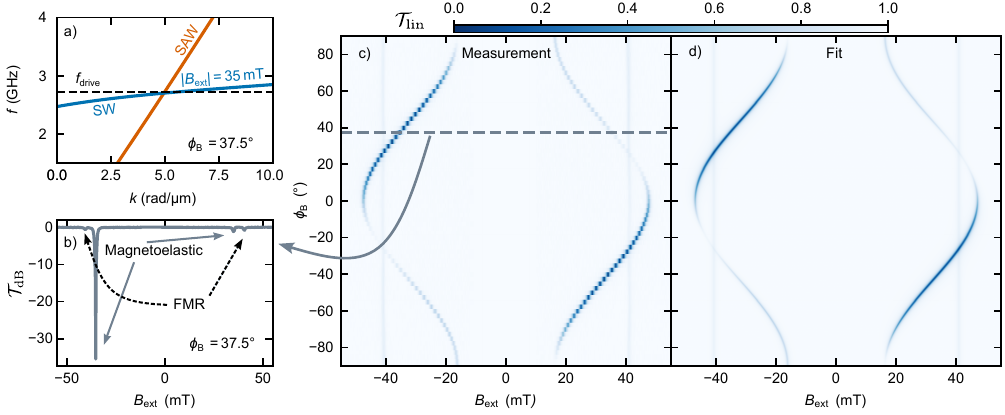}
    \caption{\label{fig:magelast} Magnetoelastic coupling in the device with $t_{\mathrm{AlScN}}=\qty{54.5}{\nano\meter}$ and $\lambda=\qty{1.2}{\micro\meter}$. a) Calculated uncoupled SAW (orange) and SW (blue) dispersions for $|B_{\mathrm{ext}}|=\qty{35}{\milli\tesla}$ and $\phi_\mathrm{B}=\qty{37.5}{\degree}$. The horizontal dashed line marks $f_{\mathrm{drive}}=\qty{2.720}{\giga\hertz}$. b) Magnetic-field dependence of the normalized transmission at $\phi_\mathrm{B}=\qty{37.5}{\degree}$. The solid gray arrows mark the magnetoelastic resonances, whereas the dashed black arrows indicate weak directly excited FMR features. c) Measured normalized transmission amplitude as a function of $B_{\mathrm{ext}}$ and $\phi_\mathrm{B}$. The horizontal dashed line indicates the field cut shown in b). d) Corresponding fit obtained from the magnetoelastic-absorption model (Eq.~(\ref{eq:t_calc})).}
\end{figure*}

Figure~\ref{fig:magelast}~a) shows the calculated uncoupled SAW and SW dispersion relations. Resonant conversion requires the conservation of both frequency and wave vector. At $k=2\pi/\lambda$, the two branches intersect close to $f_{\mathrm{drive}}$, predicting magnetoelastic coupling near $|B_{\mathrm{ext}}|=\qty{35}{\milli\tesla}$.

At fixed $f_{\mathrm{drive}}$, we define the normalized linear transmission amplitude and its logarithmic representation as:

\begin{equation}
    \label{eq:T_lin}
    \mathcal{T}_\mathrm{lin} = |S_\mathrm{21} (B_\mathrm{ext})| / |S_\mathrm{21, BL}|~,
\end{equation}

\begin{equation}
    \label{eq:T_dB}
    \mathcal{T}_\mathrm{dB} = 20 \mathrm{log_{10}} \left( \frac{|S_\mathrm{21}  (B_\mathrm{ext})|}{|S_\mathrm{21, BL}|} \right)~.
\end{equation}

Here, $S_{21,\mathrm{BL}}$ denotes the baseline transmission measured at a far-detuned magnetic field for which no magnetoelastic resonance occurs.

As shown in Fig.~\ref{fig:magelast}~b), pronounced transmission minima occur at $B_{\mathrm{ext}}=\qty{\pm35.26}{\milli\tesla}$, in agreement with the calculated resonance condition. At these fields, acoustic energy is transferred to the spin system, thereby reducing the SAW amplitude detected by the receiving IDT. The attenuation is substantially stronger for negative than for positive fields. This field-polarity asymmetry results from the helicity mismatch between the elliptical polarization of the Rayleigh-like SAW and the sense of magnetization precession \cite{Kus2021-lr}. The weaker features result from a direct, non-acoustic microwave excitation of the ferromagnetic resonance (FMR) in the device and have been observed in related magnetoacoustic structures \cite{Kunz2025-mf, Geilen2022-fd, Hatanaka2025-nb}. The maximum observed attenuation in our devices approaches $\qty{70}{\decibel\per\milli\meter}$.

The full angular dependence is shown in Fig.~\ref{fig:magelast}~c), where $\phi_\mathrm{B}$ is varied between $\qty{-90}{\degree}$ and $\qty{90}{\degree}$. The resonance-field trajectories are determined primarily by the angle-dependent SW dispersion, whereas the resonance amplitudes reflect the symmetry and relative phases of the magnetoelastic driving fields. We model this response using the effective driving field approach introduced by Dreher \textit{et al.} and extended by Küß \textit{et al.} to finite-wave-vector SWs \cite{Dreher2012-lw, Kus2020-fu}. Because the SAW propagates along the YIG $[\bar{1}10]$ direction, the cubic magnetoelastic energy and the resulting driving fields are transformed into the corresponding device coordinate system. A summary of the approach can be found in supplement Sec.~S6. The calculated power transmission is then given by:

\begin{eqnarray}
    \label{eq:t_calc}
    \mathcal{T}_\mathrm{calc}^\mathrm{p} &&=  1-\frac{P_\mathrm{abs}}{P_\mathrm{0}} \nonumber \\
    &&= \mathrm{exp} \left( - \mu_0 C f_\mathrm{drive} \int_\mathrm{V}\mathrm{Im}(\textbf{h}^\dagger_\mathrm{[\bar{1}10]}\underline{\chi}\textbf{h}_\mathrm{[\bar{1}10]})\mathrm{dV} \right)
\end{eqnarray}

Here, $P_{\mathrm{abs}}$ is the absorbed power, $P_0$ is the incident SAW power, $\underline{\chi}$ is the dynamic Polder susceptibility tensor, and $\mathbf{h}_\mathrm{[\bar{1}10]}$ is the magnetoelastic driving field detailed in Sec.~\ref{sec:realtive} in Eqs.~(\ref{eq:hip},\ref{eq:hoop}). The field-independent parameter $C$ accounts for the unknown absolute strain amplitude. Because Figs.~\ref{fig:magelast}~c) and \ref{fig:magelast}~d) show the transmission amplitude and not power, the quantity compared with the experiment is $\sqrt{\mathcal{T}_\mathrm{calc}^\mathrm{p}} \triangleq \mathcal{T}_\mathrm{lin} $. A weak phenomenological FMR contribution is additionally included to reproduce the directly excited FMR. The resulting fit, shown in Fig.~\ref{fig:magelast}~d), captures the resonance-field trajectories, angular intensity variation, and pronounced field-polarity asymmetry observed experimentally. Note that the FMR is almost angle-independent, demonstrating that the pronounced angular dependency of the magneto-elastically driven spin wave resonance stems from the anisotropic spin-wave dispersion and not from anisotropy of the YIG. In addition, from the fit we obtain an effective damping parameter of $\alpha_{\mathrm{eff}}=\alpha_{\mathrm{FMR}}+\gamma \mu_0 \Delta H_\mathrm{inhomo}/f=0.0038$, with the gyromagnetic ratio $\gamma$, the conventional gilbert damping parameter $\alpha_{\mathrm{FMR}}$ and the inhomogeneous broadening $\mu_0 \Delta H_\mathrm{inhomo}$. The obtained value is approximately half that reported for comparable YIG/ZnO devices with the same YIG thickness \cite{Kunz2025-mf}.

Because the unknown absolute strain amplitude is absorbed into the scaling parameter $C$, only the relative magnitudes and phases of the complex strain components can be extracted from the fit. The detailed description of the fitting procedure and used parameters are provided in the supplement Sec.~S6. Following Ref.~\cite{Kunz2025-mf}, the fitting model approximates both the acoustic strain and the lowest-order SW mode profile as uniform across the \qty{100}{\nano\meter}-thick YIG layer. The FEM strain fields generally retain a finite depth dependence. To map them onto the effective strain components entering the uniform model, we coherently average each complex FEM strain component over the YIG thickness. The resulting FEM and fitted strain vectors are then normalized using the same norm:

\begin{eqnarray}
    \label{eq:strain_norm}
    \varepsilon_i^\mathrm{YIG}
    &&=
    \frac{1}{t_\mathrm{YIG}}
    \int_{t_\mathrm{YIG}}
    \varepsilon_i(z)\mathrm{d}z~,  \nonumber  \\
    \varepsilon_i^{\mathrm{YIG,norm}}
    &&=
    \frac{\varepsilon_i^\mathrm{YIG}}
    {\left(
    \displaystyle\sum_{j\in\mathcal{J}}
    \left|\varepsilon_j^\mathrm{YIG}\right|^2
    \right)^{1/2}}~,
\end{eqnarray}

with $\mathcal{J}={xx,yy,xy,xz,yz}$. Here, $t_\mathrm{YIG}$ denotes the YIG thickness. The component $\varepsilon_{zz}$ is excluded from $\mathcal{J}$ because it does not contribute to the transverse magnetoelastic driving field in the present in-plane configuration and therefore cannot be constrained by the fit. We further label the relative pase between the dominant strain components $\varepsilon_{xx}$ and $\varepsilon_{xz}$ as $\vartheta_{xx-xz}$.
The resulting normalized magnitudes and relative phases are summarized in Table~\ref{tab:strain_comparison}.

\begin{table}[ht]
    \caption{\label{tab:strain_comparison}
    Comparison of the normalized magnitudes of the YIG-averaged complex strain components and the relative phase $\vartheta_{xx-xz}$ obtained from the FEM simulation and the fit to the experimental data.}
    \centering
    \begin{ruledtabular}
        \begin{tabular}{
            l
            S[table-format=2.2]
            S[table-format=2.2]
        }
            Quantity & {FEM-simulation} & {Fit} \\
            \hline
            $| \varepsilon_{xx}^\mathrm{YIG, norm}|$ & 0.71  & 0.78 \\
            $| \varepsilon_{yy}^\mathrm{YIG, norm}|$ & 0     & 0.03 \\
            $| \varepsilon_{xy}^\mathrm{YIG, norm}|$ & 0.11  & 0.05 \\
            $| \varepsilon_{xz}^\mathrm{YIG, norm}|$ & 0.66  & 0.62 \\
            $| \varepsilon_{yz}^\mathrm{YIG, norm}|$ & 0.03  & 0.01 \\
            \hline
            $\vartheta_{xx-xz}$ (\unit{\degree}) & 89.99 & 93.29 \\
        \end{tabular}
    \end{ruledtabular}
\end{table}

Both approaches identify the longitudinal component $\varepsilon_{xx}$ and the vertical-shear component $\varepsilon_{xz}$ as dominant, whereas all remaining components are small. The FEM simulation yields an almost exact quadrature relation, $\vartheta_{xx-xz}=\qty{89.99}{\degree}$, while the fit gives $\vartheta_{xx-xz}=\qty{93.29}{\degree}$ and thus remains close to quadrature. The agreement in both relative magnitude and phase is consistent with predominantly sagittal, elliptically polarized motion. Together with the surface confinement established above, it therefore supports the identification of the excited mode as Rayleigh-like.

\section{Thickness dependent transmission}\label{sec:realtive}

We next compare the relative transmission defined in Eqs.~(\ref{eq:T_lin},\ref{eq:T_dB}) for different SAW wavelengths and AlScN thicknesses. For this comparison, we select an external-field angle of $\phi_\mathrm{B}=\qty{35}{\degree}$, at which pronounced magnetoelastic attenuation is observed. Because $\mathcal{T}_\mathrm{lin}$ is a relative transmission-amplitude ratio, lower values correspond to stronger attenuation. At a fixed wavelength, varying $t_\mathrm{AlScN}$ redistributes the acoustic strain across the heterostructure and thereby changes the fraction of the strain overlapping with the YIG layer. This modifies the magnetoelastic driving fields and, consequently, the efficiency with which the SAW excites spin waves.

Following the notation introduced by Dreher \textit{et al.} \cite{Dreher2012-lw}, we calculate the two driving-field components transverse to the equilibrium magnetization for the SAW propagation along $x\parallel[\bar{1}10]$ (supplement Sec.~S6). These components are oriented within the film plane (ip) and perpendicular to the film plane (oop), respectively:

\begin{widetext}
\begin{equation}
    \label{eq:hip}
    \mu_0 h_\mathrm{[\bar{1}10]}^\mathrm{ip} = \left[ \bar{b}(\varepsilon_\mathrm{xx}-\varepsilon_\mathrm{yy}) -2\sqrt{2}\delta b\varepsilon_\mathrm{yz} \right] \mathrm{sin}(2\phi_0)
    - 2 \left[ \bar{b}\varepsilon_\mathrm{xy} - \sqrt{2}\delta b \varepsilon_\mathrm{xz}\right] \mathrm{cos}(2\phi_0)
\end{equation}

\begin{equation}
    \label{eq:hoop}
    \mu_0 h_\mathrm{[\bar{1}10]}^\mathrm{oop} = 2\left[ (\bar{b}+\delta b) \varepsilon_\mathrm{xz} - \sqrt{2}\delta b \varepsilon_\mathrm{xy} \right] \mathrm{cos}(\phi_0)
    + \left[ 2(\bar{b}+\delta b) \varepsilon_\mathrm{yz} + \sqrt{2} \delta b (\varepsilon_\mathrm{yy} - \varepsilon_\mathrm{xx}) \right] \mathrm{sin}(\phi_0)~.
\end{equation}
\end{widetext}

Here, $\phi_0$ denotes the equilibrium magnetization angle. We use the substitutions $\bar{b}=(b_\mathrm{1}+2b_\mathrm{2})/3$ and $\delta b=(b_\mathrm{1}-b_\mathrm{2})/3$ to simplify the expressions. The energy-density magnetoelastic coefficients of YIG are $B_\mathrm{1}=\qty{3.48e5}{\joule\per\meter\cubed}$ and $B_\mathrm{2}=\qty{6.96e5}{\joule\per\meter\cubed}$ \cite{Comstock1965-gz}. These are converted to field-normalized convention, $b_i=B_i/M_\mathrm{s}$ with saturation magnetization $M_\mathrm{s}$ of YIG.

The FEM simulations provide the complex, depth-dependent displacement and strain profiles of each acoustic eigenmode. Because the overall amplitude of an eigenvector returned by the eigensolver is arbitrary, the raw strain amplitudes cannot be compared directly between different modes. We therefore normalize each mode by the $L^2$ norm of its displacement profile and use the complex strain averaged over the YIG layer, as defined in Eq.~(\ref{eq:strain_norm}). Introducing the displacement normalization factor:
\begin{equation}
    \mathcal{N}_u =
    \left[
    \int_\mathrm{All}
    \mathbf{u}^\dagger(z)\mathbf{u}(z)\mathrm{d}z
    \right]^{1/2},
    \qquad
    \varepsilon_i^\mathrm{eff}
    =
    \frac{\varepsilon_i^\mathrm{YIG}}{\mathcal{N}_u},
    \label{eq:epseff}
\end{equation}
provides a common normalization for all eigenmodes. Here, the index $i$ runs over the six independent strain components, and the integration extends over the complete simulated layer stack. Because strain depends linearly on displacement, this normalization removes the arbitrary eigenvector amplitude. The same real-valued normalization factor is applied to all strain components of a given mode, thereby preserving their relative amplitudes and phases. Only the physically irrelevant global phase of the eigenmode remains arbitrary.

The resulting $\varepsilon_i^\mathrm{eff}$ represents the coherently averaged strain within the YIG layer under a common displacement-normalization convention. It therefore allows the relative trends in $\mathcal{T}_\mathrm{calc}^\mathrm{p}$ to be compared consistently between eigenmodes with different wavelengths and AlScN thicknesses. However, this normalization does not assign the same acoustic energy or power to the different modes. Consequently, $\varepsilon_i^\mathrm{eff}$, the corresponding magnetoelastic driving fields, and the calculated transmission remain relative quantities. 

Averaging the complex strain over the YIG thickness automatically accounts for phase variations within the layer through coherent cancellation. The remaining approximation is that the magnetoelastic overlap with the lowest-order spin-wave mode can be represented by this thickness-averaged strain. This treatment is reasonable for the present \qty{100}{\nano\meter}-thick YIG film, which corresponds to only \qty{3.6}{\percent}-\qty{8.3}{\percent} of the investigated acoustic wavelengths.

\begin{figure}
\includegraphics{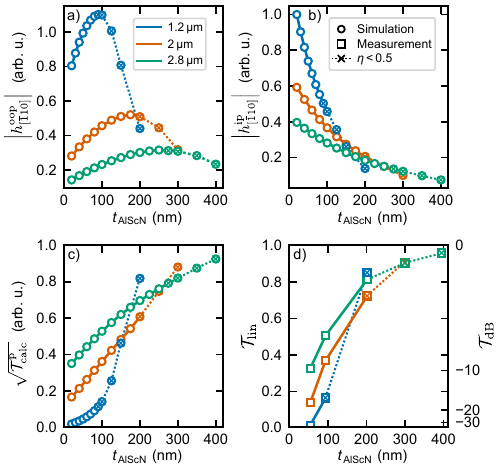}
\caption{\label{fig:calc} Dependence of $|h_\mathrm{[\bar{1}10]}^\mathrm{oop, ip}|$ and normalized transmission on the $t_\mathrm{AlScN}$. Calculated magnitudes of the a) $|h_\mathrm{[\bar{1}10]}^\mathrm{ip}|$ and b) $|h_\mathrm{[\bar{1}10]}^\mathrm{oop}|$. c) Calculated normalized amplitude-transmission $\sqrt{\mathcal{T}_\mathrm{calc}^\mathrm{p}}$, obtained using a common value of $C$ for all wavelengths and thicknesses. d) Measured normalized amplitude transmission $\mathcal{T}_\mathrm{lin}$, with the corresponding decibel scale $\mathcal{T}_\mathrm{dB}$ shown on the right axis. Colors denote the nominal acoustic wavelength. Circles and squares represent simulations and measurements, respectively, while crosses identify modes with $\eta<0.5$.}
\end{figure}

Figure~\ref{fig:calc}~a,b) shows the resulting magnitudes of the relative oop and ip driving-field components. The colors denote the wavelengths, following the convention of Fig.~\ref{fig:freq}: \qty{1.2}{\micro\meter} in blue, \qty{2}{\micro\meter} in orange, and \qty{2.8}{\micro\meter} in green. Simulated data are represented by circles, whereas the measurements in Fig.~\ref{fig:calc}~d) are represented by squares. Crosses identify modes with $\eta<0.5$, that are not strictly surface localized.

For the Rayleigh-like modes considered here, the two driving-field components are dominated by different parts of the strain field: $|h_\mathrm{oop}|$ primarily follows $\varepsilon_\mathrm{xz}^\mathrm{eff}$, whereas $|h_\mathrm{ip}|$ is governed mainly by $\varepsilon_\mathrm{xx}^\mathrm{eff}$. This distinction explains their different dependencies on $t_\mathrm{AlScN}$. The shear strain $\varepsilon_\mathrm{xz}$ exhibits a maximum magnitude below the free surface. Increasing $t_\mathrm{AlScN}$ therefore initially improves its overlap with the YIG layer, before the overlap decreases again as the mode progressively loses surface confinement. This produces the nonmonotonic behavior of $|h_\mathrm{oop}|$. By contrast, $\varepsilon_\mathrm{xx}$ is largest near the free surface and decreases with depth, leading to an approximately monotonic reduction of $|h_\mathrm{ip}|$ with increasing $t_\mathrm{AlScN}$. 

Although the magnitudes shown in Fig.~\ref{fig:calc}~a,b) provide a useful qualitative picture, they do not fully determine the magnetoelastic attenuation. Both driving-field components are complex, and the calculation in Eq.~(\ref{eq:t_calc}) uses the complete quantity $\mathbf{h}^\dagger_\mathrm{[\bar{1}10]}\underline{\chi}\mathbf{h}_\mathrm{[\bar{1}10]}$. It therefore accounts for both the relative phase between the driving-field components and the complex response described by the Polder susceptibility tensor. The magnitude-based interpretation given above should consequently be understood as an intuitive explanation of the main trends rather than a complete description of the calculated attenuation.

Using Eq.~(\ref{eq:t_calc}), together with the driving fields calculated above and a single scaling factor $C$ for all modes, we calculate the field-dependent relative transmission. The Polder susceptibility tensor entering this expression depends strongly on the external magnetic field. When the field satisfies the SAW-SW resonance condition, the dissipative response obtained from its contraction with the magnetoelastic driving fields becomes maximal, resulting in the strongest predicted attenuation. Accordingly, for each wavelength and AlScN thickness, $\sqrt{\mathcal{T}_\mathrm{calc}^\mathrm{p}}$ is evaluated at the resonance condition. Because the absolute strain amplitude associated with a simulated eigenmode is arbitrary, $C$ cannot be determined from the eigenmode calculation alone. Consequently, $\sqrt{\mathcal{T}_\mathrm{calc}^\mathrm{p}}$ should not be interpreted as an absolute device transmission. Nevertheless, using the same value of $C$ places all modes on a common relative scale and allows their thickness- and wavelength-dependent trends to be compared consistently.

As shown in Fig.~\ref{fig:calc}~c), $\sqrt{\mathcal{T}_\mathrm{calc}^\mathrm{p}}$ is enhanced with increasing $t_\mathrm{AlScN}$ for all investigated wavelengths, indicating a reduction in the relative magnetoelastic attenuation. Within the strongly surface-localized regime, the shortest-wavelength and therefore highest-frequency mode exhibits the strongest predicted attenuation. The origin of this behavior is twofold. First, as shown in panels a) and b), the effective strain components within the YIG and the resulting magnetoelastic driving fields are largest for the shortest-wavelength mode. Second, Eq.~(\ref{eq:t_calc}) contains an explicit frequency dependence such that, for otherwise fixed parameters, $\sqrt{\mathcal{T}_\mathrm{calc}^\mathrm{p}}$ decreases with increasing $f_\mathrm{drive}$. Because the shorter wavelengths correspond to higher drive frequencies in the present device series, this frequency dependence further enhances their predicted attenuation.

As $t_\mathrm{AlScN}$ increases, the wavelength-dependent loss of surface confinement changes the relative ordering of the transmission branches. In particular, the shortest-wavelength mode loses strong surface localization at a smaller AlScN thickness than the longer-wavelength modes. Its strain overlap with the YIG layer therefore begins to decrease at a smaller thickness, weakening the magnetoelastic driving fields and reducing the predicted attenuation. Consequently, the corresponding transmission branch rises and eventually intersects the branches of the longer-wavelength modes.

The measurements in Fig.~\ref{fig:calc}~d) qualitatively reproduce these principal trends. For each wavelength and AlScN thickness, the transmission data are first time-gated to suppress electromagnetic crosstalk. The magnetic-field-dependent normalized transmission is then evaluated at the fixed external-field angle $\phi_\mathrm{B}=\qty{35}{\degree}$, matching the angle used in the calculations. The minimum value of $\mathcal{T}_\mathrm{lin}$ is extracted from a narrow magnetic-field sweep around the expected resonance field and plotted in Fig.~\ref{fig:calc}~d). This minimum corresponds to the maximum measured magnetoelastic attenuation. For each wavelength series, $\mathcal{T}_\mathrm{lin}$ increases with for thicker AlScN, confirming that the relative attenuation is enhanced in devices with thinner AlScN layers. At comparable thicknesses within the thin-film regime, the \qty{1.2}{\micro\meter} devices exhibit the lowest normalized transmission and hence the strongest relative attenuation, in agreement with the calculation.

A direct quantitative prediction of the transmission is beyond the present model because it would require an energy- or power-normalized acoustic mode and the complete coherent, depth-dependent overlap between the acoustic strain and the spin-wave profile. Nevertheless, for the \qty{100}{\nano\meter}-thick YIG layer and the lowest-order spin-wave mode considered here, the thickness-averaged approximation provides a reasonable basis for comparing the relative trends.

The calculation additionally predicts a flattening of the \qty{1.2}{\micro\meter} transmission branch for $t_\mathrm{AlScN}\lesssim\qty{100}{\nano\meter}$, suggesting that further reductions in the piezoelectric-layer thickness produce only a limited additional change in the relative transmission. This behavior is not resolved experimentally because the thickness range below approximately \qty{100}{\nano\meter} is sampled too sparsely. Additional devices with closely spaced AlScN thicknesses in this regime would be required to test the predicted saturation.

\subsection{Absolute transmission}\label{subsec:absolute}

\begin{figure}[ht]
\includegraphics[width=\columnwidth]{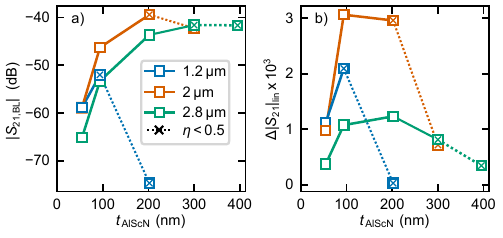}
\caption{\label{fig:absolut}
a) Baseline transmission magnitude $|S_{21,\mathrm{BL}}|$ (off resonant with spin system), expressed in decibels. b) Absolute transmission change $\Delta |S_{21}|_{\mathrm{lin}}$, in linear amplitude units. Crossed symbols denote modes with $\eta<0.5$, which are not classified as strongly surface localized. Lines are guides to the eye.}
\end{figure}

Having established the thickness dependence of the normalized transmission, we finally consider the absolute change in the measured transmission. A large relative attenuation does not necessarily produce a large detectable signal when the off-resonant SAW transmission is already weak. Figure~\ref{fig:absolut}~a) shows the time-gated baseline transmission magnitude $|S_{21,\mathrm{BL}}|$, where the acoustic and spin-wave modes are off resonance, expressed in decibels. This is the same baseline used to normalize the transmission data above. For the IDT geometries employed in this work, the baseline transmission of the $t_{\mathrm{AlScN}}=\qty{54.5}{\nano\meter}$ devices is lower than that of the $t_{\mathrm{AlScN}}=\qty{94}{\nano\meter}$ devices for all three wavelengths.

The baseline transmission is governed by the complete electroacoustic response of the delay line. In the present layered structure, the effective electromechanical coupling coefficient $K^2$ depends on the normalized AlScN thickness $t_{\mathrm{AlScN}}/\lambda$. For sufficiently thin piezoelectric layers, $K^2$ can become small, thereby reducing the acoustic generation and detection efficiencies \cite{Smith1969-mf, Wang2014-xz}. Conversely, increasing the AlScN thickness can reduce the surface confinement and promote substrate radiation, which further decreases the received signal \cite{FARNELL197235, Lefeuvre1998-ct, Lefeuvre1998-ht}. The AlScN thickness also affects the electrical capacitance and acoustic loading of the IDT and therefore its input impedance and matching to the microwave circuitry \cite{Kino1973-ri}. Consequently, an IDT geometry optimized for a particular AlScN thickness may no longer provide optimal electrical matching when the layer thickness is changed.

To compare the magnitude of the magnetoelastic transmission feature directly, we define the positive resonance-induced change in the linear $S_\mathrm{21}$ magnitude as:

\begin{equation}
    \label{eq:dS21}
    \Delta |S_\mathrm{21}|_\mathrm{lin} = |S_\mathrm{21, BL}| - |S_\mathrm{21}(B_\mathrm{res})|~,
\end{equation}

where $B_{\mathrm{res}}$ denotes the field at which the magnetoelastic transmission minimum occurs. All quantities in Eq.~(\ref{eq:dS21}) are evaluated after time gating.

As shown in Fig.~\ref{fig:absolut}~b), the largest measured $\Delta|S_{21}|_{\mathrm{lin}}$ occurs at $t_{\mathrm{AlScN}}=\qty{94}{\nano\meter}$ for $\lambda=\qty{1.2}{\micro\meter}$ and $\qty{2}{\micro\meter}$. For $\lambda=\qty{2.8}{\micro\meter}$, the maximum occurs near $t_{\mathrm{AlScN}}=\qty{200}{\nano\meter}$, although the value at $\qty{94}{\nano\meter}$ remains comparable. Thus, for the wavelengths and IDT geometries investigated here, an AlScN thickness of approximately $\qty{100}{\nano\meter}$ provides a practical compromise between relative magnetoelastic attenuation and absolute detectable signal. Accordingly, this value provides a useful reference point for comparable magnetoelastic delay-line devices, particularly when detailed optimization of the IDT geometry is impractical, but should not be interpreted as a universal optimum.

\section{Conclusion}

In conclusion, we have demonstrated magnetoacoustic coupling in AlScN/YIG/GGG heterostructures and shown that the AlScN thickness provides an effective means of controlling the interaction between surface acoustic waves and spin waves. Cross-sectional microscopy reveals crystalline AlScN on YIG, separated by an amorphous interfacial layer of only approximately \qty{3.5}{\nano\meter}, highlighting the suitability of AlScN for direct integration with YIG. The interfacial properties could be further optimized in future devices by adjusting deposition parameters, such as the growth temperature and plasma conditions, or by introducing a thin dielectric buffer layer, for example $\mathrm{SiO_2}$. Field- and angle-resolved transmission measurements reveal magnetoacoustic attenuation approaching \qty{70}{\decibel\per\milli\meter}, while the extracted effective damping is lower than that obtained in comparable ZnO/YIG devices. By combining the angular dependence of the magnetoacoustic response with finite-element simulations, we identify the excited modes as predominantly Rayleigh-like and relate the measured coupling to the complex strain field within the YIG layer. The simulations further reveal that surface confinement in this fast-on-slow material stack depends strongly on both AlScN thickness and acoustic wavelength. Reducing the AlScN thickness enhances the strain overlap with YIG and therefore the relative magnetoelastic attenuation. The largest absolute change in the transmitted signal, however, occurs near an AlScN thickness of \qty{100}{\nano\meter} for the present devices, reflecting the competing influences of magnetoelastic coupling, electromechanical transduction, acoustic leakage, and IDT matching. This optimum is consequently device- and wavelength-dependent. Our results establish AlScN/YIG as a promising magnetoacoustic platform and provide design considerations for balancing baseline SAW transmission with efficient SAW-SW coupling in future devices.

\begin{acknowledgments}
This work was supported by the European Research Council (ERC) under the European Union’s Horizon Europe research and innovation programme (Consolidator Grant ”MAWiCS”, Grant Agreement No. 101044526), the Deutsche Forschungsgemeinschaft (DFG, German Research Foundation) within the Transregional Collaborative Research Center TRR 173/3-268565370 Spin+X (Project B13).
Microstructuring took place at the Nano Structuring Center (NSC) at the RPTU Kaisersalutern. XRD and TEM was carried out at the IFOS Kaiserslautern.
\end{acknowledgments}

\section*{Author contributions}
K.K. and M.W. conceived the project. K.K., A.{\v{Z}}. and S.B. fabricated the samples. K.K. and K.G. carried out the measurements. K.K. analyzed and interpreted the results in discussion with all coauthors. K.K. wrote the manuscript with input from all coauthors.

\section*{Competing interests}
The authors declare that they have no competing interests.

\section*{Data availability}
The data supporting the findings of this study is available on Zenodo \cite{ZenodoData}.

\bibliographystyle{apsrev4-2}
\bibliography{Bibliography}

@article{Kuenstle2025,
  author  = {K{\"u}nstle, Kevin and Kunz, Yannik and Moussa, Tarek and Lasinger, Katharina and Yamamoto, Kei and Pirro, Philipp and Gregg, John F. and Kamra, Akashdeep and Weiler, Mathias},
  title   = {Magnon-polaron control in a surface magnetoacoustic wave resonator},
  journal = {Nat. Commun.},
  volume  = {16},
  pages   = {10116},
  year    = {2025},
  doi     = {10.1038/s41467-025-66301-x}
}

@article{Yang2021-yz,
  author  = {Yang, Wei-Gang and Schmidt, Holger},
  title   = {Acoustic control of magnetism toward energy-efficient applications},
  journal = {Appl. Phys. Rev.},
  volume  = {8},
  pages   = {021304},
  year    = {2021},
  doi     = {10.1063/5.0042138}
}

@article{Krenner2026-wi,
  author  = {Krenner, Hubert J. and Santos, Paulo V. and Westerhausen, Christoph and Andersson, Gustav and Cleland, Andrew N. and Sellier, Hermann and Takada, Shintaro and B{\"a}uerle, Christopher and Wigger, Daniel and Kuhn, Tilmann and Machnikowski, Pawe{\l} and Wei{\ss}, Matthias and Moody, Galan and Hern{\'a}ndez-M{\'\i}nguez, Alberto and Lazi{\'c}, Sne{\v{z}}ana and Kuznetsov, Alexander S. and K{\"u}{\ss}, Matthias and Albrecht, Manfred and Weiler, Mathias and Puebla, Jorge and Hwang, Yunyoung and Otani, Yoshichika and Balram, Krishna C. and Chen, I-Tung and Lai, Keji and Li, Mo and Nash, Geoff R. and Nysten, Emeline D. S. and Bhattacharjee, Paromita and Mishra, Himakshi and Iyer, Parameswar K. and Nemade, Harshal B. and Khelif, Abdelkrim and Benchabane, Sarah and Feng, Gao and Jin, Yabin and Bartasyte, Ausrine and Margueron, Samuel and Marangolo, Massimiliano and Thevenard, Laura and Rovillain, Pauline and Gourdon, Catherine and Hage-Ali, Sami and Elmazria, Omar and Schmidt, Hagen and Yeo, Leslie Y. and Ambattu, Lizebona A. and Jeon, Jessie S. and Kwak, Daesik and Rufo, Joseph and Yang, Shujie and Huang, Tony Jun},
  title   = {The 2026 guided acoustic waves roadmap},
  journal = {J. Phys. D: Appl. Phys.},
  volume  = {59},
  pages   = {093001},
  year    = {2026},
  doi     = {10.1088/1361-6463/ae258d}
}

@article{Puebla2022-ja,
  author  = {Puebla, Jorge and Hwang, Yunyoung and Maekawa, Sadamichi and Otani, Yoshichika},
  title   = {Perspectives on spintronics with surface acoustic waves},
  journal = {Appl. Phys. Lett.},
  volume  = {120},
  pages   = {220502},
  year    = {2022},
  doi     = {10.1063/5.0093654}
}

@book{royer1999elastic,
  author     = {Royer, Daniel and Dieulesaint, Eug{\`e}ne},
  translator = {Lyle, Stephen N.},
  title      = {Elastic Waves in Solids {II}: Generation, Acousto-optic Interaction, Applications},
  series     = {Advanced Texts in Physics},
  publisher  = {Springer-Verlag},
  address    = {Berlin},
  year       = {2000},
  doi        = {10.1007/978-3-662-06938-7}
}

@article{Thevenard2016-ou,
  author  = {Thevenard, L. and Camara, I. S. and Majrab, S. and Bernard, M. and Rovillain, P. and Lema{\^\i}tre, A. and Gourdon, C. and Duquesne, J.-Y.},
  title   = {Precessional magnetization switching by a surface acoustic wave},
  journal = {Phys. Rev. B},
  volume  = {93},
  pages   = {134430},
  year    = {2016},
  doi     = {10.1103/PhysRevB.93.134430}
}

@article{Yang2024-rd,
  author  = {Yang, Yang and Zhao, Le and Yi, Di and Xu, Teng and Chai, Yahong and Zhang, Chenye and Jiang, Dingsong and Ji, Yahui and Hou, Dazhi and Jiang, Wanjun and Tang, Jianshi and Yu, Pu and Wu, Huaqiang and Nan, Tianxiang},
  title   = {Acoustic-driven magnetic skyrmion motion},
  journal = {Nat. Commun.},
  volume  = {15},
  pages   = {1018},
  year    = {2024},
  doi     = {10.1038/s41467-024-45316-w}
}

@article{Schwenke2025-gg,
  author  = {Schwenke, Philipp and Spindler, Ephraim and Vasyuchka, Vitaliy I. and Hamadeh, Alexandre Abbass and Pirro, Philipp and Weiler, Mathias},
  title   = {Ratchet motion of magnetic skyrmions driven by surface acoustic sawtooth waves},
  journal = {Phys. Rev. B},
  volume  = {112},
  pages   = {214409},
  year    = {2025},
  doi     = {10.1103/w75m-hlsr}
}

@article{Kus2023-wh,
  author  = {K{\"u}{\ss}, Matthias and Hassan, M. and Kunz, Yannik and H{\"o}rner, Andreas and Weiler, Mathias and Albrecht, Manfred},
  title   = {Nonreciprocal transmission of magnetoacoustic waves in compensated synthetic antiferromagnets},
  journal = {Phys. Rev. B},
  volume  = {107},
  pages   = {214412},
  year    = {2023},
  doi     = {10.1103/PhysRevB.107.214412}
}

@article{Kus2021-lr,
  author  = {K{\"u}{\ss}, Matthias and Heigl, Michael and Flacke, Luis and Hefele, Andreas and H{\"o}rner, Andreas and Weiler, Mathias and Albrecht, Manfred and Wixforth, Achim},
  title   = {Symmetry of the magnetoelastic interaction of {Rayleigh} and shear horizontal magnetoacoustic waves in nickel thin films on {$\mathrm{LiTaO}_3$}},
  journal = {Phys. Rev. Applied},
  volume  = {15},
  pages   = {034046},
  year    = {2021},
  doi     = {10.1103/PhysRevApplied.15.034046}
}

@article{Hwang2024-ks,
  author  = {Hwang, Yunyoung and Puebla, Jorge and Kondou, Kouta and Gonzalez-Ballestero, Carlos and Isshiki, Hironari and S{\'a}nchez Mu{\~n}oz, Carlos and Liao, Liyang and Chen, Fa and Luo, Wei and Maekawa, Sadamichi and Otani, Yoshichika},
  title   = {Strongly coupled spin waves and surface acoustic waves at room temperature},
  journal = {Phys. Rev. Lett.},
  volume  = {132},
  pages   = {056704},
  year    = {2024},
  doi     = {10.1103/PhysRevLett.132.056704}
}

@article{Alekseev2025-dk,
  author  = {Alekseev, S. G. and Polzikova, N. I. and Luzanov, V. A. and Raevskiy, A. O. and Nikitov, S. A.},
  title   = {Linear and parametric phonon-magnon interactions in hybrid magnetoacoustic resonator},
  journal = {J. Appl. Phys.},
  volume  = {138},
  pages   = {023901},
  year    = {2025},
  doi     = {10.1063/5.0275317}
}

@misc{Kunstle2026-bn,
  author        = {K{\"u}nstle, Kevin and Wagner, Matthias and Knaus, Philipp and Kunz, Yannik and Spindler, Ephraim and Lasinger, Katharina and Schweizer, Matthias R. and Pirro, Philipp and Gregg, John F. and Weiler, Mathias},
  title         = {Nonlinear frequency shift and bistability of magnon-polarons},
  year          = {2026},
  eprint        = {2605.22157},
  archiveprefix = {arXiv},
  primaryclass  = {cond-mat.mes-hall},
  doi           = {10.48550/arXiv.2605.22157}
}

@article{CHEREPANOV199381,
  author  = {Cherepanov, Vladimir and Kolokolov, Igor and L'vov, Victor},
  title   = {The saga of {YIG}: Spectra, thermodynamics, interaction and relaxation of magnons in a complex magnet},
  journal = {Phys. Rep.},
  volume  = {229},
  pages   = {81--144},
  year    = {1993},
  doi     = {10.1016/0370-1573(93)90107-O}
}

@article{Ba2025-dz,
  author  = {Ba, You and Puebla, Jorge and Yamamoto, Kei and Hwang, Yunyoung and Liao, Liyang and Maekawa, Sadamichi and Klein, Olivier and Otani, Yoshichika},
  title   = {Nonreciprocal resonant surface acoustic wave absorption in {$\mathrm{Y}_3\mathrm{Fe}_5\mathrm{O}_{12}$}},
  journal = {Phys. Rev. B},
  volume  = {111},
  pages   = {104401},
  year    = {2025},
  doi     = {10.1103/PhysRevB.111.104401}
}

@article{Holzmann2025-mv,
  author  = {Holzmann, Christian and K{\"u}{\ss}, Matthias and Glamsch, Stephan and Stein, David and Kunz, Yannik and Weiler, Mathias and Albrecht, Manfred},
  title   = {Polycrystalline {YIG} thin films on a piezoelectric substrate for magnetoacoustic hybrid devices},
  journal = {ACS Appl. Mater. Interfaces},
  volume  = {17},
  pages   = {58550--58558},
  year    = {2025},
  doi     = {10.1021/acsami.5c14928}
}

@article{Ryburn2025-fq,
  author  = {Ryburn, Finlay and K{\"u}nstle, Kevin and Zhang, Yangzhan and Kunz, Yannik and Reimann, Timmy and Lindner, Morris and Dubs, Carsten and Gregg, John F. and Weiler, Mathias},
  title   = {Generation of gigahertz-frequency surface acoustic waves in {$\mathrm{Y}_3\mathrm{Fe}_5\mathrm{O}_{12}/\mathrm{ZnO}$} heterostructures},
  journal = {Phys. Rev. Applied},
  volume  = {23},
  pages   = {034062},
  year    = {2025},
  doi     = {10.1103/PhysRevApplied.23.034062}
}

@article{Hatanaka2025-nb,
  author  = {Hatanaka, Daiki and Asano, Motoki and Kurosu, Megumi and Taniyasu, Yoshitaka and Okamoto, Hajime and Yamaguchi, Hiroshi},
  title   = {On-chip magnon-polaron generation in mode-matched cavity magnomechanics},
  journal = {Phys. Rev. Applied},
  volume  = {24},
  pages   = {064053},
  year    = {2025},
  doi     = {10.1103/l84d-4tgb}
}

@article{Rivard2025-je,
  author  = {Rivard, Carson and Jander, Albrecht and Dhagat, Pallavi},
  title   = {Sustained amplification of coherent spin waves by parametric pumping with surface acoustic waves},
  journal = {npj Spintronics},
  volume  = {3},
  pages   = {49},
  year    = {2025},
  doi     = {10.1038/s44306-025-00115-x}
}

@incollection{FARNELL197235,
  author    = {Farnell, G. W. and Adler, E. L.},
  title     = {Elastic Wave Propagation in Thin Layers},
  booktitle = {Physical Acoustics},
  editor    = {Mason, Warren P. and Thurston, R. N.},
  volume    = {9},
  pages     = {35--127},
  publisher = {Academic Press},
  address   = {New York},
  year      = {1972},
  doi       = {10.1016/B978-0-12-395670-5.50007-6}
}

@article{Kurz2019-nz,
  author  = {Kurz, Nicolas and Ding, Anli and Urban, Daniel F. and Lu, Yuan and Kirste, Lutz and Feil, Niclas M. and {\v{Z}}ukauskait{\.e}, Agn{\.e} and Ambacher, Oliver},
  title   = {Experimental determination of the electro-acoustic properties of thin film {AlScN} using surface acoustic wave resonators},
  journal = {J. Appl. Phys.},
  volume  = {126},
  pages   = {075106},
  year    = {2019},
  doi     = {10.1063/1.5094611}
}

@article{Urban2021-pg,
  author  = {Urban, Daniel F. and Ambacher, Oliver and Els{\"a}sser, Christian},
  title   = {First-principles calculation of electroacoustic properties of wurtzite {(Al,Sc)N}},
  journal = {Phys. Rev. B},
  volume  = {103},
  pages   = {115204},
  year    = {2021},
  doi     = {10.1103/PhysRevB.103.115204}
}

@article{Zukauskaite2012-yt,
  author  = {{\v{Z}}ukauskait{\.e}, Agn{\.e} and Wingqvist, Gunilla and Palisaitis, Justinas and Jensen, Jens and Persson, Per O. {\AA}. and Matloub, Ramin and Muralt, Paul and Kim, Yunseok and Birch, Jens and Hultman, Lars},
  title   = {Microstructure and dielectric properties of piezoelectric magnetron sputtered {$w$-Sc$_x$Al$_{1-x}$N} thin films},
  journal = {J. Appl. Phys.},
  volume  = {111},
  pages   = {093527},
  year    = {2012},
  doi     = {10.1063/1.4714220}
}

@article{Zywitzki2017-nu,
  author  = {Zywitzki, Olaf and Modes, Thomas and Barth, Stephan and Bartzsch, Hagen and Frach, Peter},
  title   = {Effect of scandium content on structure and piezoelectric properties of {AlScN} films deposited by reactive pulse magnetron sputtering},
  journal = {Surf. Coat. Technol.},
  volume  = {309},
  pages   = {417--422},
  year    = {2017},
  doi     = {10.1016/j.surfcoat.2016.11.083}
}

@article{Dreher2012-lw,
  author  = {Dreher, L. and Weiler, M. and Pernpeintner, M. and Huebl, H. and Gross, R. and Brandt, M. S. and Goennenwein, S. T. B.},
  title   = {Surface acoustic wave driven ferromagnetic resonance in nickel thin films: Theory and experiment},
  journal = {Phys. Rev. B},
  volume  = {86},
  pages   = {134415},
  year    = {2012},
  doi     = {10.1103/PhysRevB.86.134415}
}

@article{Kunz2025-mf,
  author  = {Kunz, Yannik and Sch{\"u}ler, Julian and Ryburn, Finlay and K{\"u}nstle, Kevin and Schneider, Michael and Lasinger, Katharina and Zhang, Yangzhan and Pirro, Philipp and Gregg, John F. and Weiler, Mathias},
  title   = {Efficient spin-wave excitation by surface acoustic waves in ultralow-damping yttrium iron garnet--zinc oxide heterostructures},
  journal = {Phys. Rev. Applied},
  volume  = {24},
  pages   = {014043},
  year    = {2025},
  doi     = {10.1103/97kd-ysh6}
}

@article{Osterlund2021-hk,
  author  = {{\"O}sterlund, Elmeri and Ross, Glenn and Caro, Miguel A. and Paulasto-Kr{\"o}ckel, Mervi and Hollmann, Andreas and Klaus, Manuela and Meixner, Matthias and Genzel, Christoph and Koppinen, Panu and Pensala, Tuomas and {\v{Z}}ukauskait{\.e}, Agn{\.e} and Trebala, Michal},
  title   = {Stability and residual stresses of sputtered wurtzite {AlScN} thin films},
  journal = {Phys. Rev. Materials},
  volume  = {5},
  pages   = {035001},
  year    = {2021},
  doi     = {10.1103/PhysRevMaterials.5.035001}
}

@incollection{Pupalaikis_Doshi_2013,
  author    = {Pupalaikis, Peter J. and Doshi, Kaviyesh},
  title     = {{TDR}-based {S}-parameters},
  booktitle = {Modern {RF} and Microwave Measurement Techniques},
  editor    = {Teppati, Valeria and Ferrero, Andrea and Sayed, Mohamed},
  pages     = {279--306},
  publisher = {Cambridge University Press},
  address   = {Cambridge},
  year      = {2013},
  doi       = {10.1017/CBO9781139567626.012}
}

@misc{COMSOLMultiphysics,
  author = {{COMSOL AB}},
  title  = {{COMSOL Multiphysics}},
  year   = {2020},
  note   = {Version 5.6 (build 5.6.0.280), Stockholm, Sweden},
  url    = {https://www.comsol.com}
}

@article{Lefeuvre1998-ct,
  author  = {Lefeuvre, O. and Zinin, P. and Briggs, G. A. D. and Every, A.},
  title   = {Surface wave dispersion beyond cutoff for a fast layer on a slow substrate},
  journal = {Appl. Phys. Lett.},
  volume  = {72},
  pages   = {856--857},
  year    = {1998},
  doi     = {10.1063/1.120916}
}

@article{Lefeuvre1998-ht,
  author  = {Lefeuvre, O. and Zinin, P. and Briggs, G. A. D.},
  title   = {Leaky surface waves propagating on a fast on slow system and the implications for material characterization},
  journal = {Ultrasonics},
  volume  = {36},
  pages   = {229--232},
  year    = {1998},
  doi     = {10.1016/S0041-624X(97)00058-9}
}

@article{Wang2014-xz,
  author  = {Wang, Wenbo and Mayrhofer, Patrick M. and He, Xingli and Gillinger, Manuel and Ye, Zhi and Wang, Xiaozhi and Bittner, Achim and Schmid, Ulrich and Luo, J. K.},
  title   = {High performance {AlScN} thin film based surface acoustic wave devices with large electromechanical coupling coefficient},
  journal = {Appl. Phys. Lett.},
  volume  = {105},
  pages   = {133502},
  year    = {2014},
  doi     = {10.1063/1.4896853}
}

@article{Smith1969-mf,
  author  = {Smith, W. R. and Gerard, H. M. and Collins, J. H. and Reeder, T. M. and Shaw, H. J.},
  title   = {Analysis of interdigital surface wave transducers by use of an equivalent circuit model},
  journal = {IEEE Trans. Microw. Theory Tech.},
  volume  = {17},
  pages   = {856--864},
  year    = {1969},
  doi     = {10.1109/TMTT.1969.1127075}
}

@article{Yamamoto2022-sk,
  author  = {Yamamoto, Kei and Xu, Mingran and Puebla, Jorge and Otani, Yoshichika and Maekawa, Sadamichi},
  title   = {Interaction between surface acoustic waves and spin waves in a ferromagnetic thin film},
  journal = {J. Magn. Magn. Mater.},
  volume  = {545},
  pages   = {168672},
  year    = {2022},
  doi     = {10.1016/j.jmmm.2021.168672}
}

@article{Geilen2022-fd,
  author  = {Geilen, Moritz and Nicoloiu, Alexandra and Narducci, Daniele and Mohseni, Morteza and Bechberger, Moritz and Ender, Milan and Ciubotaru, Florin and Hillebrands, Burkard and M{\"u}ller, Alexandru and Adelmann, Christoph and Pirro, Philipp},
  title   = {Fully resonant magneto-elastic spin-wave excitation by surface acoustic waves under conservation of energy and linear momentum},
  journal = {Appl. Phys. Lett.},
  volume  = {120},
  pages   = {242404},
  year    = {2022},
  doi     = {10.1063/5.0088924}
}

@article{Kus2020-fu,
  author  = {K{\"u}{\ss}, Matthias and Heigl, Michael and Flacke, Luis and H{\"o}rner, Andreas and Weiler, Mathias and Albrecht, Manfred and Wixforth, Achim},
  title   = {Nonreciprocal {Dzyaloshinskii--Moriya} magnetoacoustic waves},
  journal = {Phys. Rev. Lett.},
  volume  = {125},
  pages   = {217203},
  year    = {2020},
  doi     = {10.1103/PhysRevLett.125.217203}
}

@article{Comstock1965-gz,
  author  = {Comstock, R. L.},
  title   = {Magnetoelastic coupling constants of the ferrites and garnets},
  journal = {Proc. IEEE},
  volume  = {53},
  pages   = {1508--1517},
  year    = {1965},
  doi     = {10.1109/PROC.1965.4263}
}

@article{Kino1973-ri,
  author  = {Kino, G. S. and Wagers, R. S.},
  title   = {Theory of interdigital couplers on nonpiezoelectric substrates},
  journal = {J. Appl. Phys.},
  volume  = {44},
  pages   = {1480--1488},
  year    = {1973},
  doi     = {10.1063/1.1662397}
}

@article{Davis2010,
  author  = {Davis, S. and Baruth, A. and Adenwalla, S.},
  title   = {Magnetization dynamics triggered by surface acoustic waves},
  journal = {Appl. Phys. Lett.},
  volume  = {97},
  pages   = {232507},
  year    = {2010},
  doi     = {10.1063/1.3521289}
}

@ARTICLE{Serga2010YIGMagnonics,
  author  = {Serga, Alexander A. and Chumak, Andrii V.
             and Hillebrands, Burkard},
  title   = {{YIG} magnonics},
  journal = {J. Phys. D: Appl. Phys.},
  volume  = {43},
  number  = {26},
  pages   = {264002},
  year    = {2010},
  doi     = {10.1088/0022-3727/43/26/264002}
}

@ARTICLE{Dubs2017LPEYIG,
  author  = {Dubs, Carsten and Surzhenko, Oleksii and Linke, Ralf
             and Danilewsky, Andreas and Br{\"u}ckner, Uwe and Dellith, Jan},
  title   = {Sub-micrometer yttrium iron garnet {LPE} films with low
             ferromagnetic resonance losses},
  journal = {J. Phys. D: Appl. Phys.},
  volume  = {50},
  number  = {20},
  pages   = {204005},
  year    = {2017},
  doi     = {10.1088/1361-6463/aa6b1c}
}

@BOOK{Pozar2012MicrowaveEngineering,
  author    = {Pozar, David M.},
  title     = {Microwave Engineering},
  edition   = {4th},
  publisher = {John Wiley \& Sons},
  address   = {Hoboken, NJ},
  year      = {2012},
  isbn      = {9780470631553}
}

@ARTICLE{Darinskii1997LeakyWaves,
  author  = {Darinskii, A. N.},
  title   = {On the theory of leaky waves in crystals},
  journal = {Wave Motion},
  volume  = {25},
  number  = {1},
  pages   = {35--49},
  year    = {1997},
  doi     = {10.1016/S0165-2125(96)00031-5}
}

@ARTICLE{Laude2004SAWSlowness,
  author  = {Laude, Vincent and Masson, Mathias and Ballandras, Sylvain
             and Solal, Marc},
  title   = {Imaginary branches of surface acoustic wave slowness curves},
  journal = {J. Appl. Phys.},
  volume  = {96},
  number  = {11},
  pages   = {6895--6902},
  year    = {2004},
  doi     = {10.1063/1.1813639}
}

@INPROCEEDINGS{Naumenko1998LeakyLayered,
  author    = {Naumenko, N. F. and Didenko, I. S.},
  title     = {Leaky wave propagation in layered structures},
  booktitle = {Proceedings of the 1998 IEEE Ultrasonics Symposium},
  volume    = {1},
  pages     = {149--154},
  publisher = {IEEE},
  year      = {1998},
  doi       = {10.1109/ULTSYM.1998.762120}
}

@ARTICLE{Every2018FanoLeakySAW,
  author  = {Every, A. G. and Maznev, A. A.},
  title   = {{Fano} line shapes of leaky surface acoustic waves extending
             from supersonic surface wave points},
  journal = {Wave Motion},
  volume  = {79},
  pages   = {1--9},
  year    = {2018},
  doi     = {10.1016/j.wavemoti.2018.02.005}
}

@article{Barth2022,
  author  = {Barth, Stephan and Schreiber, Tom and Cornelius, Steffen and Zywitzki, Olaf and Modes, Thomas and Bartzsch, Hagen},
  title   = {High Rate Deposition of Piezoelectric {AlScN} Films by Reactive Magnetron Sputtering from {AlSc} Alloy Targets on Large Area},
  journal = {Micromachines},
  volume  = {13},
  number  = {10},
  pages   = {1561},
  year    = {2022},
  doi     = {10.3390/mi13101561}
}

@article{Schuetz2015,
  author  = {Schuetz, M. J. A. and Kessler, E. M. and Giedke, G. and Vandersypen, L. M. K. and Lukin, M. D. and Cirac, J. I.},
  title   = {Universal Quantum Transducers Based on Surface Acoustic Waves},
  journal = {Phys. Rev. X},
  volume  = {5},
  number  = {3},
  pages   = {031031},
  year    = {2015},
  doi     = {10.1103/PhysRevX.5.031031}
}

@misc{ZenodoData,
  author    = {K{\"u}nstle, K. and Gauch, K. and {\v{Z}}ukauskait{\.e}, A. and Barth, S. and Weiler, M.},
  title     = {Data supporting this study},
  year      = {2026},
  publisher = {Zenodo},
  doi       = {10.5281/zenodo.23687273},
  note      = {Available at \url{https://doi.org/10.5281/zenodo.23687273}}
}

\end{document}